\documentclass[manuscript]{emulateapj}

\usepackage{graphicx}
\usepackage{dcolumn}
\usepackage{bm}
\usepackage{xcolor}
\usepackage{color}
\def\MagEff{$-0.0072(37)$}
\def\MagEffS{$0.0237$}
\def\WrongHost{$+0.0007(09)$}                        \def\WrongHostS{$0.0059$}   
\def\PBA{$-0.0109(03)$}
\def\PBAS{$0.0192$}
\def\DJones{$-0.0028(14)$}
\def\DJonesS{$0.0089$}
\def\IAX{$-0.0022(09)$}
\def\IAXS{$0.0059$}
\def\MagEffData{$-0.006$}
\def\MagEffsig{0.05$\sigma$}

\def\PBAData{$-0.041$}
\def\PBAsig{1.57$\sigma$}
\def\DJonesData{$-0.002$}
\def\DJonessig{0.09$\sigma$}
\def\IAXData{$+0.001$}
\def\IAXsig{2.07$\sigma$}

\def\MagEffp{$-0.0072 \pm  0.0037$}

\def\CCcontamination{4}

\def\CCKTen{1}

\def\CCIAX{5.9}

\usepackage[hang]{footmisc}

\begin{document}

\preprint{APS/123-QED}

\title{Assessment of systematic uncertainties in the cosmological analysis of the SDSS supernovae photometric sample}

\submitted{}

\author{Brodie Popovic\footnotemark[1], Dan Scolnic\footnotemark[1], Richard Kessler\footnotemark[2]$^,$\footnotemark[3]}

\affiliation{$^1$Department of Physics, Duke University, Durham, NC, 27708, USA.}
\affiliation{$^2$Department of Astronomy and Astrophysics, The University of Chicago, Chicago, IL 60637, USA.} \affiliation{$^3$Kavli Institute for Cosmological Physics, University of Chicago, Chicago, IL 60637, USA.}

\date{\today}

\begin{abstract}
Improvement in the precision of measurements of cosmological parameters with Type Ia Supernovae (SNIa) is expected to come from large photometrically identified (photometric) SN samples.  Here we re-analyse the SDSS photometric SN sample, with roughly 700 high-quality, likely but unconfirmed SNIa light-curves, to develop new analysis tools aimed at evaluating systematic uncertainties on the dark energy equation-of-state parameter $w$. Since we require a spectroscopically measured host galaxy redshift for each SN, we determine the associated selection efficiency of host galaxies in order to simulate bias corrections.  We determine that the mis-association rate of host galaxies is $0.6\%$; ignoring this effect in simulated bias corrections leads to a $w$-bias of $\Delta w = +0.0007$, where $w$ is evaluated from SNIa and priors from measurements of baryon acoustic oscillations and the cosmic microwave background. We assess the uncertainty in our modeling of the host galaxy selection efficiency and find the associated $w$ uncertainty to be $-0.0072$.  Finally, we explore new core collapse (CC) models in simulated training samples and find that adjusting the CC luminosity distribution to be in agreement with previous Pan-STARRS analyses yields a better match to the SDSS data. The impact of ignoring this adjustment is $\Delta w = -0.0109$; the impact of replacing the new CC models with those used by Pan-STARRS is $\Delta w = -0.0028$. These systematic uncertainties are subdominant to the statistical constraints from the SDSS sample, but must be considered in future photometric analyses of large SN samples such as those from DES, LSST and WFIRST.
\end{abstract}

\maketitle

\section{\label{sec:level1}Introduction}

\footnotetext[1]{Email: brodie.popovic@duke.edu}

\newcommand{\SNhostas}{\Delta\theta}
\newcommand{\Trest}{T_{\rm{rest}}}
\newcommand{\MJDpeak}{\rm{MJD}_{\rm{peak}}}
\newcommand{\DDLRratio}{$r_{\textrm{\tiny{DLR}}}$}
\newcommand{\stellarm}{M_{\rm{stellar}}}
\newcommand{\effHost}{\epsilon_{\rm host}(r)}
\newcommand{\effz}{\epsilon_{\rm host}(z)}

The discovery of accelerating cosmic expansion \citep{Riess98, Perlmutter99} from measurements of 10s of Type Ia supernovae (SNIa) galvanized a new era in the study of cosmology. In the time since this discovery, collections of 100s of spectroscopically confirmed SNIa have been used to measure the expansion history of the universe up to $z = 1$ \citep{Conley11, Betoule14, Scolnic18, DES3YR}. A combination of constraints from SNIa and those from other probes such as baryon acoustic oscillations (BAO; \citealp{Eisenstein05, Anderson13}) and the cosmic microwave background (CMB; \citealp{Bennett03, Planck15}) can be used to infer the dark energy equation-of-state parameter $w = P/\rho c^2$ where $P$ is the pressure and $\rho$ is the energy density. The most precise of these measurements is that of \cite{Scolnic18} (hereafter S18) which used 1,048 spectroscopically confirmed SNIa. Measurements of SNIa, combined with constraints from \cite{Planck16}, yield \textit{w} = $-1.026 \pm 0.041$.

Significant improvement in the constraint on dark energy from supernovae requires a large jump in the supernova sample size. Unfortunately, obtaining such a large number of spectroscopic confirmations for SNe is unfeasible with expected resources in the next decade. Time constraints limit single-object spectroscopy, and the sparse density of supernovae (~10 yr$^{-1}$ deg$^{-2}$ with \textit{R}-band magnitude $<$ 22) makes the yield for multi-object spectroscopy similarly low. On the other hand, spectroscopic classification may not be necessary if one can use photometric classification of the light-curve sample. The difficulty with photometric analysis is that it is susceptible to contamination from core collapse (CC) SNe and possible contamination from peculiar SNIa and non-supernova transients such as AGN \citep{Campbell13, Jones18}. Significant effort has been made in classification algorithms (e.g. PSNID -  \citealp{Sako08}, SuperNNova -  \citealp{Moller19}, Nearest Neighbour - \citealp{Kessler16}, and machine learning methods -  \citealp{Lochner16}), spurred on by the advent of Pan-STARRS (PS1; \citealp{Jones18}), Dark Energy Survey (DES; \citealp{Bernstein12}), Large Synoptic Survey Telescope (LSST; \citealp{LSST}), and other SN surveys.

The first cosmological measurement of \textit{w} with primarily photometric classification was done by \cite{Campbell13}. CC contamination was reduced using the PSNID Bayesian light-curve classifier \citep{Sako11}, resulting in a final sample with 3.9\% CC SNe contamination as predicted by rigorous simulations. However, no systematic uncertainty budget was included in their analysis.

To optimally account for this contamination, \cite{Kunz07} developed the Bayesian Estimation Applied to Multiple Species (BEAMS) method to independently model the SNIa and CC Hubble residual distributions. BEAMS samples both Ia and CC species of supernovae and simultaneously fits for the contribution of each while marginalising over nuisance parameters. BEAMS relies on a classifier, such as the aforementioned PSNID or SuperNNova, to assign SNIa probabilities and bifurcate the distribution into likely Ia and CC. The first cosmological measurement using BEAMS was done by \cite{Hlozek12} on the SDSS sample but did not include a systematic uncertainty budget. The systematic uncertainties were considered by \cite{Knights13}, which developed a BEAMS formalism that gives reliable estimations of cosmological parameters. \cite{Jones18} used their own implementation of BEAMS and were the first to evaluate the systematic uncertainty budget for a photometric sample (the PS1 sample).

BEAMS was further improved by \cite{Kessler16}, incorporating bias corrections and the option of using a simulated CC sample instead of ad-hoc fit parameters to describe the CC Hubble residuals. This method, known as BEAMS with Bias Corrections (BBC), was first applied to a real photometric sample in \cite{Jones17}. The bias correction component was included in S18 for their spectroscopic sample and also the DES 3-year sample \citep{DES3YR, Brout18SYS}.

Analysing a sample with contamination relies on accurate models of CC supernovae to simulate training samples for classifiers and validation. Previous analyses have used spectroscopically confirmed light-curves of non-Ia SN to develop rest-frame Spectral Energy Distribution (SED) templates for simulations, in order to generate CC events at all redshifts. The first collection of non-Ia SED templates came from  \cite{Kessler10}, which released a simulated sample of mixed SNIa and non-Ia light-curves for a classification challenge -  Supernova Photometric Classification Challenge (SNPHOTCC). Most recently, the  Photometric LSST Astronomical Time-series Classification Challenge (PLAsTiCC; \citealp{PLAsTiCC18}) has gathered a large library of new SED templates \citep{Kessler19} that can be used for simulating training samples, further expounding upon previous efforts. The SED templates included in this release span a wider variety of transient events than in SNPHOTCC. The PLAsTiCC SED templates have not yet been used to simulate training samples as part of the analysis of a real photometric sample; here we make the first attempt. 

Current cosmological analyses with photometric samples use spectroscopically confirmed host galaxy redshifts to create a Hubble diagram. A systematic method of identifying host galaxies was introduced in \cite{Sullivan11} with the concept of directional light radius (DLR), which uses galaxy orientation and spatial size to determine the most likely host galaxy for each supernova. This DLR method was further explored in \cite{Gupta16} and \cite{Sako18}. \cite{Gupta16} include other properties in the host-assignment, and evaluate the frequency of mis-association of the host galaxies. This systematic uncertainty was evaluated in \cite{Jones17}, and is evaluated here with an improved technique. The spectroscopic targeting of galaxies based on their brightness should cause an additional systematic bias in the cosmological measurements because host galaxy properties have been found to be correlated with supernova luminosity \citep{Sullivan11}; we assess the impact of this bias. 

To examine the impact of CC modeling in simulated training sets for classifiers, as well as host galaxy selection, we perform a re-analysis of the SDSS-II Supernova Survey \citep{Frieman08}. Here, we use models from PLAsTiCC for the simulated training sample and the BBC method to construct our Hubble diagram. This paper is the first of two works. In this work, we evaluate systematic uncertainties in the photometric analysis of the SDSS sample that would not be included in a conventional spectroscopic analysis, e.g. S18. This paper also includes a broader range of CC models using PLAsTiCC, and improved methods for evaluating systematic uncertainties arising from mis-associated hosts. SDSS is the only publicly available photometric sample without previously applied selection cuts. Therefore, methods presented here can be checked and improved by the community.

In the next paper, we will measure nuisance and cosmological parameters from this sample and compare to those from the Pan-STARRS photometric sample. We will also combine these two photometric samples for a cosmological measurement.

The layout of this paper is as follows. A review of the data is in section 2. Analysis techniques and assignment of host galaxies are in section 3. The simulations for bias corrections and training samples are described in section 4. An evaluation of different systematic uncertainties is explored in section 5. Finally, the conclusions are in section 6. 

\section{Data Sample}\label{sec:level2}

The Sloan Digital Sky Survey began in 2000 as the first wide-area sky survey using charge-coupled devices \citep{York2000}. 
A review of the SDSS Supernova Survey is given in \cite{Frieman08}. In brief, Stripe 82 (from Right Ascension of 20$^h$ to 04$^h$ and 2.5$^{\textrm{o}}$ wide along the equator in Declination) was repeatedly scanned every four days using \textit{ugriz} filters. The processing pipeline for the images is described in \cite{Stoughton02} and potential SNe were identified in subtracted images with the method developed by \cite{Alard98}. Candidate selection and spectroscopic identification are described by \cite{Sako08}. The photometry is described in \cite{Holtzman08}. The spectroscopically confirmed subset of this data was used in several analyses to measure cosmological parameters  \citep{Kessler09, Betoule14, Scolnic18}.

Through three observing seasons (Fall 2005 through Fall 2007), the SDSS supernova program discovered 10,258 new variable objects \citep{Sako18} and measured their \textit{ugriz} light-curves. Further specifics of the supernovae population statistics can be found within \cite{Sako18}. A component of the SDSS supernova survey included spectroscopic follow-up for a limited number of identified host galaxies.

\begin{figure}[h]
\includegraphics[width=8cm]{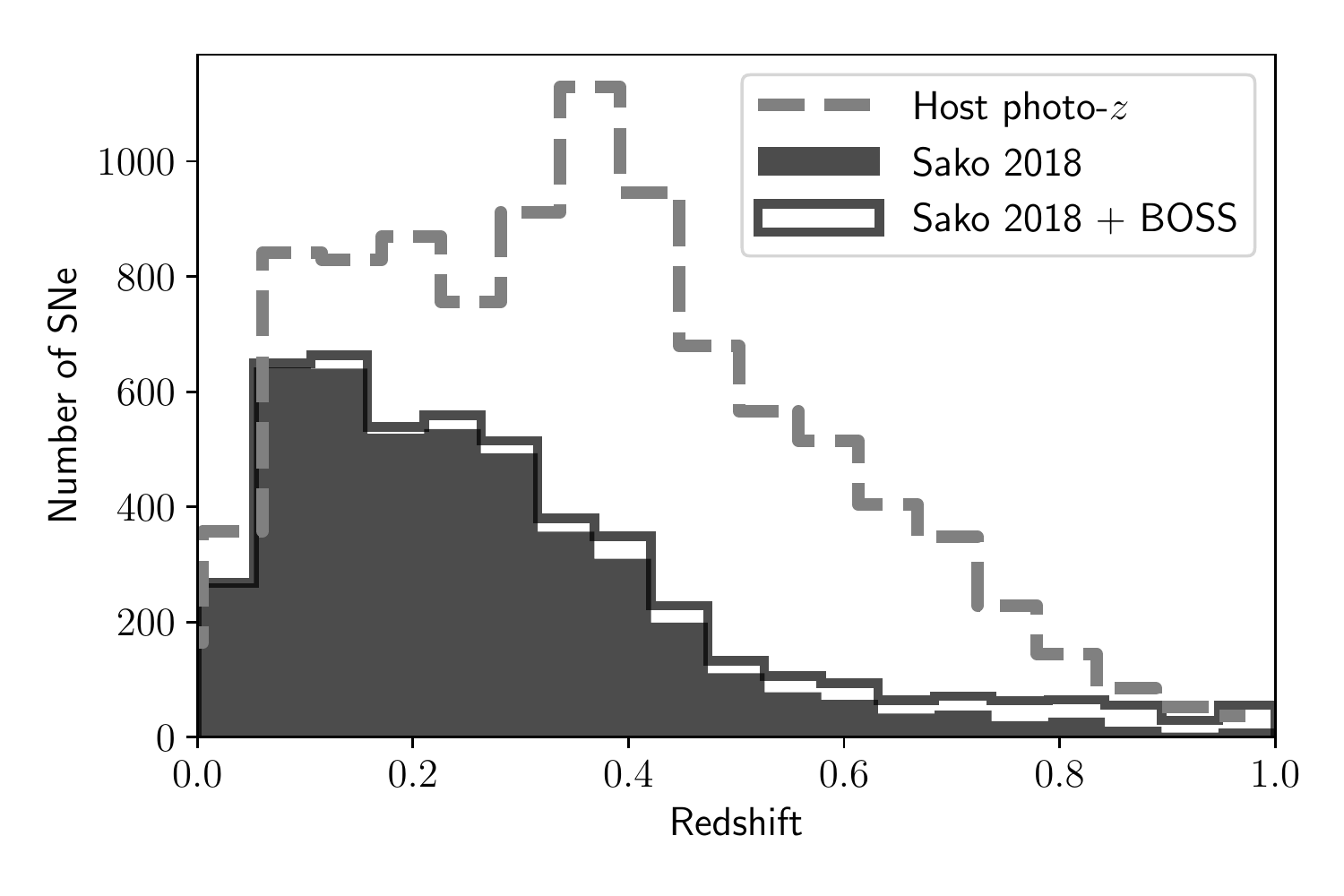}
\caption{The redshift distribution of the SDSS transient sample. In filled histogram we show the distribution of spec-$z$ of the host galaxies from Sako 2018; in solid histogram we show the distribution including the 1,294 host galaxy spec-$z$ from BOSS. The dashed histogram is the host galaxy photo-$z$ distribution found for all SDSS transients.}
\label{zCMB-Hist}
\end{figure}

A separate SDSS spectroscopic survey (the Baryon Oscillation Spectroscopic Survey, or BOSS) acquired a significant fraction of potential host galaxy redshifts \citep{Dawson13}. For 4,680 candidates, they obtained an accurate spectroscopic redshift of the corresponding host galaxy. Since the conclusion of the SDSS Supernova Survey in 2008, BOSS has acquired an additional 1,294 host galaxy spectroscopic redshifts. Figure \ref{zCMB-Hist} shows the spectroscopic redshift (spec-$z$) and photometric redshift (photo-$z$) distributions for SDSS and BOSS.

The data release from \cite{Sako18} has galaxies assigned using the position of the SNe from the first SN detection epoch, which typically has low signal-to-noise (SNR); here we re-assign the host galaxies using the averaged position of the SNe from all detections. This change in coordinate calculation is minor - the mean difference in angular separation values is 0.1 arcseconds with a standard deviation of 1.4 arcseconds.

\section{Analysis}\label{sec:level3}
Measuring cosmological parameters with SNIa requires modeling and fitting the observed light-curves and using the results to standardize supernova brightness for distance measurements. Distances are evaluated with the Tripp equation, detailed in Equation 4 in \cite{DES3YR} and Equation 3 in S18. Simulations are also needed to correct for distance biases and to generate training samples for classifiers. We review the main steps of the analysis here. 

To better understand our systematic uncertainties, we simulated 30,000 events and split them into 40 data sized samples so that each simulated subsample is comparable in size to our real data sample ($\sim$700 SNe). The full analysis is performed on each simulated subsample as well as the true data sample. 

\subsection{Host Matching}
\label{sec:level3:subsec:1}

To match a supernova to its most likely host galaxy, the $d_{\textrm{\tiny{DLR}}}$ method is employed. This uses angular separation ($\SNhostas$) and accounts for the galaxy spatial profile and orientation. The derivation for $d_{\textrm{\tiny{DLR}}}$ is shown in the appendix. \cite{RaviThesis} also provides a detailed derivation. 

For each SN, all galaxies within 30$^{\prime\prime}$ are selected and sorted by ascending $d_{\textrm{\tiny{DLR}}}$ values. The galaxy with the smallest $d_{\textrm{\tiny{DLR}}}$ is considered to be the host galaxy. 

\subsection{Light-curve Fitting} \label{sec:leve2:subsec:2}

Supernova light-curve fits are done with the SALT2 light-curve model (\citealp{Guy10}, hereafter G10) using the improved model from the Joint Light-curve Analysis (JLA; \citealp{Betoule14}). The light-curve fitting and selection requirements are implemented with the SuperNova Analysis (SNANA) software package \citep{SNANA}.

Selection requirements (cuts) are applied to reduce CC contamination and to define a sample which has distance biases that can be modeled with a Monte Carlo simulation. Supernovae with properties within the SALT2 training range of colour ($ -0.3 < c < 0.3$) and stretch ($ -3 < x_1 < 3 $) are selected. To ensure well-measured light-curve fit parameters, we apply cuts on the uncertainties for stretch ($\sigma_{x_1} < 1$) and time of maximum brightness ($\sigma_{t_0} < 2$ observer frame days). We also require that the SALT2 fit probability (based on $\chi^2$ per degree of freedom) is $>$ 0.001. Next, we define a rest-frame age, $\Trest = (\rm{MJD} - \MJDpeak)/(1+$\it{z}$)$\rm, 
where MJD is the observation date, $\MJDpeak$ is the date of peak brightness, and $z$ is the redshift. We require that at least one observation satisfies $\Trest> 10$~days, and that at least one observation satisfies $\Trest < 0$~days. Finally, we require that at least two bands have an epoch in which the SNR is $> 5$. A summary of these cuts on the data are shown in Table \ref{tab:Candidates}.

\begin{table}[h]
\caption{Sequential selection requirements on the SDSS transient sample. }
\label{tab:Candidates}
\begin{tabular}{l|l|l}
Cut & Number & Comments \\
\hline
Total Candidates & 10,258 &  Data Release (Sako 2018) \\
Spec-z and $d_{\textrm{\tiny{DLR}}}$  & 4,356 & Host galaxy redshift \\
$\Trest$ $<$ 0  & 3,852 &  Light-curve sampling \\

$\Trest$ $>$ 10 & 3,518 &  Light-curve sampling\\
SNR & 2,717 & Signal to Noise Ratio $>$ 5\\
SALT2 Fit Parameters &  1,219 & $c$, $x_1$, $\sigma_{x_1}$, $\sigma_{t_0}$ \\
\hline
NN Classifier & 700 & Nearest Neighbour classifier  \\
 & &  
\end{tabular}
\end{table}

\subsection{Classification}
\label{sec:level2:subsec:5}
We use a Nearest Neighbour (NN) classifier developed by \cite{Sako18} and \cite{Kessler16}. We simulate a large training sample of SN (Ia + CC) with the same selection requirements and light-curve fitting as for the data. The redshift, colour, and stretch ($z$, $c$, and $x_1$) are used to define a 3-dimensional (3D) space for the NN analysis. In this space, each data point (real or simulated) is the center of a sphere at $\{ z, c, x_1 \}$. Classification is done by counting the number of Ia and CC SN within the sphere, and the data point is classified to be the type that is most frequent inside the sphere. The size of the sphere is set with a metric determined by maximising the product of the efficiency and purity \citep{Kessler16}. Data identified as SNIa with a probability of less than $0.5$ are rejected.

\subsection{BBC and Cosmology Fitting} \label{sec:leve2:subsec:4}
Cosmological analysis within the BBC framework is done in three stages. The first stage classifies supernovae as Ia or CC using the NN method, and assigns a probability ($P_{Ia}$) for each event to be a SNIa. The second stage separates the data into redshift bins and determines a mean distance modulus in each bin, after accounting for selection biases and CC contamination \citep{Kessler16}. Here we use 10 equal sized bins ranging from $z$ = 0.02 to 0.5. The third stage performs a cosmological fit to the binned distances using BAO \citep{Eisenstein05} and CMB \citep{Komatsu09} priors, similar to \cite{Lasker2019} who found these priors to be sufficient for systematics studies.

\begin{figure*}[t]
\includegraphics[width=18cm]{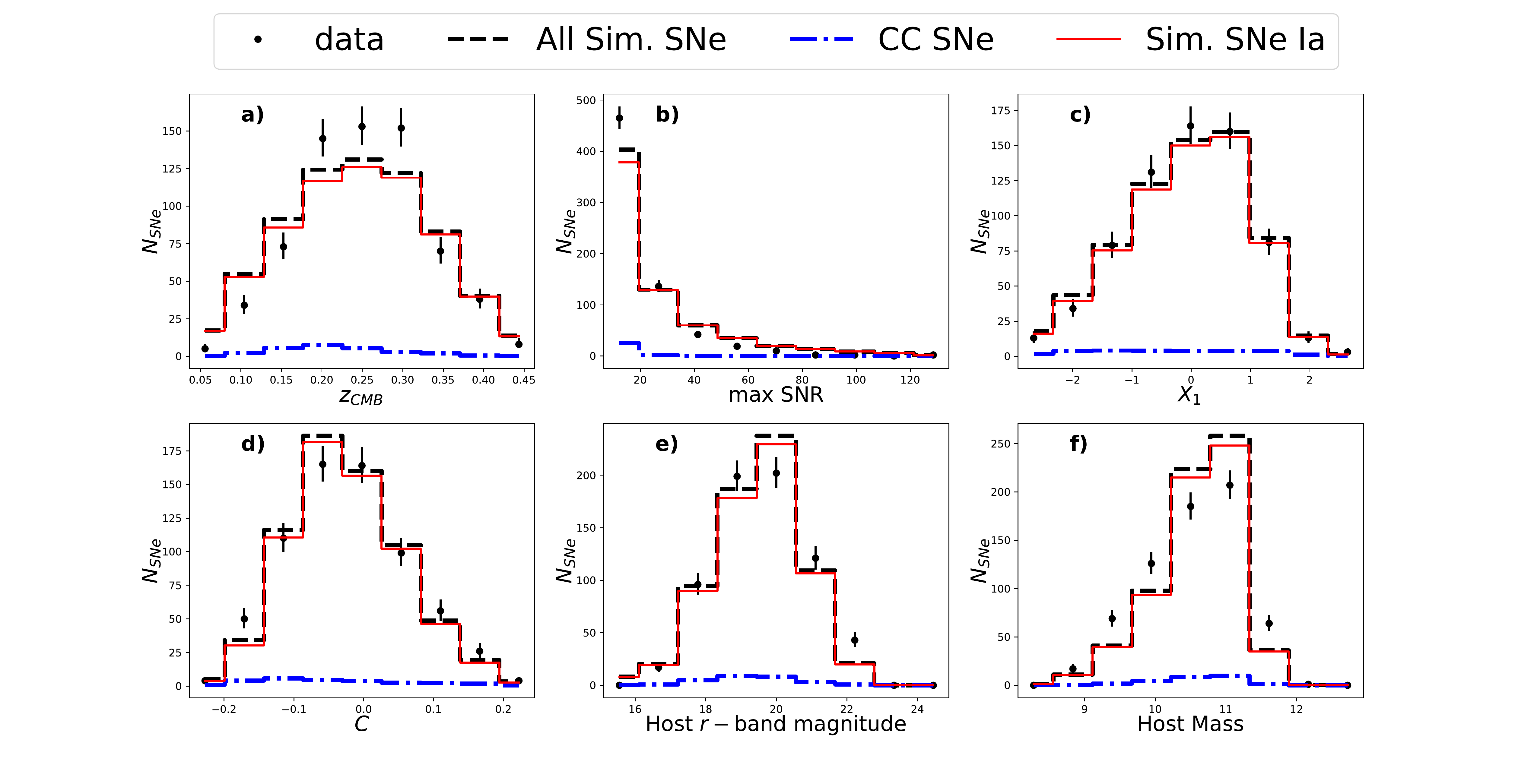}
\caption{A comparison between the data and the Fiduciary simulation for various distributions. The data is shown in points with error bars, the dashed histogram shows the total simulated SNe (Ia + CC), the solid histogram shows the simulated SNIa only, and the dotted histogram shows simulated CC. The simulated CC contamination is $\sim$\CCcontamination{}\% of the total sample and is discussed further in section \ref{sec:level5}.}
\label{simcomphist}
\end{figure*}

\begin{figure*}[t!]
\includegraphics[width=18cm]{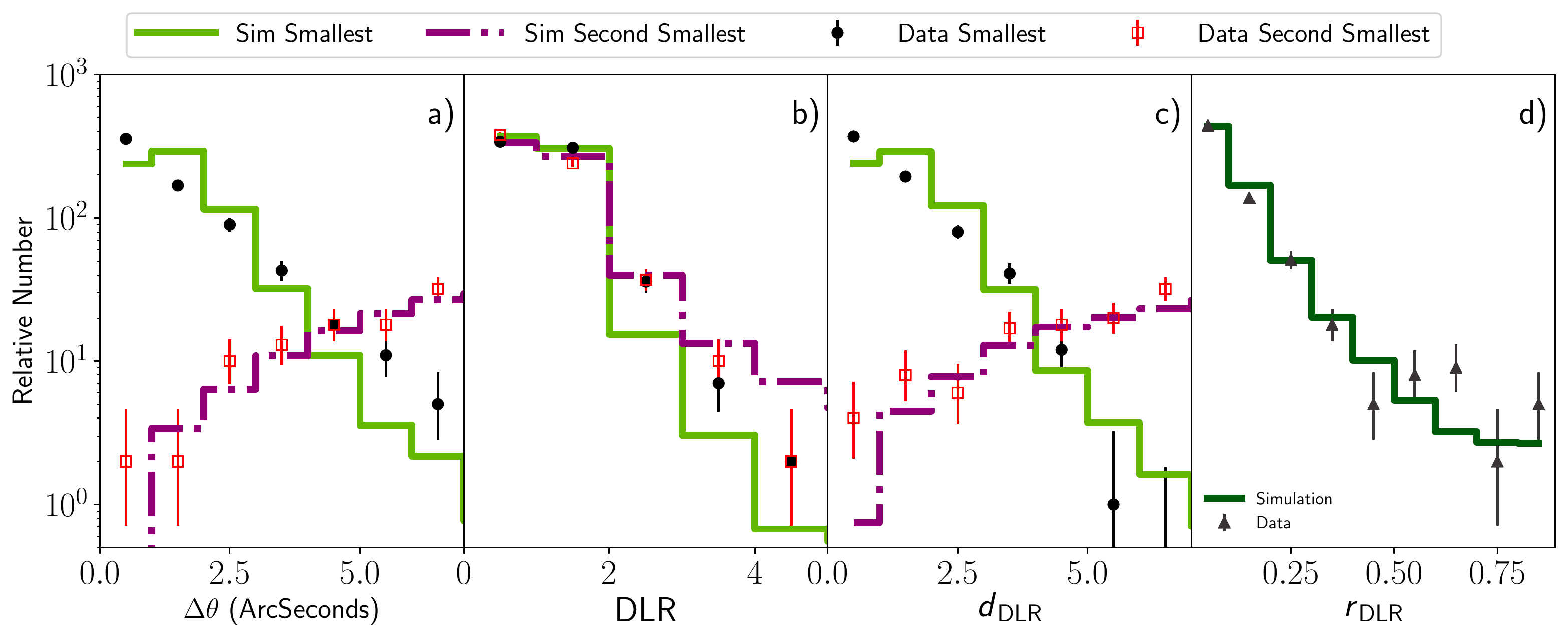}
\caption{Distributions for data (filled circles and open squares) and Fiduciary SNIa-only simulation (histogram) for quantity indicated in each panel, as defined in section \ref{sec:level3:subsec:1}. For panels a), b), and c) the filled circles and solid histogram are for the smallest $d_{\textrm{\tiny{DLR}}}$ value; open squares and dashed histogram are for the second smallest value. In d), \DDLRratio~ is shown by triangles for data and solid histogram for simulation.}
\label{DLR-3fig}
\end{figure*}

\section{Simulations} \label{sec:level4}

Simulated supernovae are needed to calculate bias corrections, create training samples, and assess the impact of our systematics. We use the SNANA software to simulate supernova light-curves using SDSS detection efficiencies, Point Spread Function (PSF) sizes, sky noise, and zeropoints. The observing history has previously been modeled for SDSS in \cite{SNANA}. While \cite{SNANA} modeled the spectroscopic selection efficiency of the supernovae, here we model the efficiency of obtaining a spectroscopic redshift from the host galaxy, as described below. A complete description of how these simulations are generated for DES is given in \cite{Kessler19}; for this study we replace the survey description of DES with that for SDSS. We simulate our SNe with $\Lambda$CDM cosmology, with a flat universe (k$= 0$), $\Omega_{\rm{Matter}} = 0.3$ and $w = -1$. 

For this analysis, we simulate both Type Ia and CC light-curves. For SNIa, we use the SALT2 model with population parameters from \cite{Scolnic16}. Simulations were generated with the stretch-luminosity $\alpha$ and colour-luminosity $\beta$ that were measured from the data as input ($\alpha = 0.14$ and $\beta$ = 3.2; \citealp{Sako18}). We use the ``G10" intrinsic scatter model from \cite{Kessler13} using the error parameterisation from \cite{Guy10}.  CC modeling is done with SED templates from \cite{Kessler10}, \cite{Jones18}, and the PLAsTiCC templates from \cite{plasticc_models} and \cite{Kessler19}, further detailed in section \ref{sec:level5:subsec:3}. We denote our `Fiduciary' analysis as using PLAsTiCC CC templates (excluding SNIax) in the training sample along with an adjusted luminosity function to match \cite{Jones17}. Figure \ref{simcomphist} shows the distributions of data are in good agreement with those from the Fiduciary simulations for redshift, SNR, colour and stretch.

\subsection{Host galaxy libraries and comparison of distributions between data and simulations} \label{sec:level3:subsec:2}

To model the potential measurement biases of cosmological parameters based on selection of host galaxies, we first create a realistic library of host galaxies (HOSTLIB) with properties that match those of our data. We evaluate the quality of our HOSTLIB by comparing the distributions of smallest and second smallest $d_{\textrm{\tiny{DLR}}}$ (section \ref{sec:level3:subsec:1}); these distributions are sensitive to galaxy spatial profile, survey depth, galaxy photo-$z$, and $\SNhostas$.

\begin{figure}[h]
\includegraphics[width=8cm]{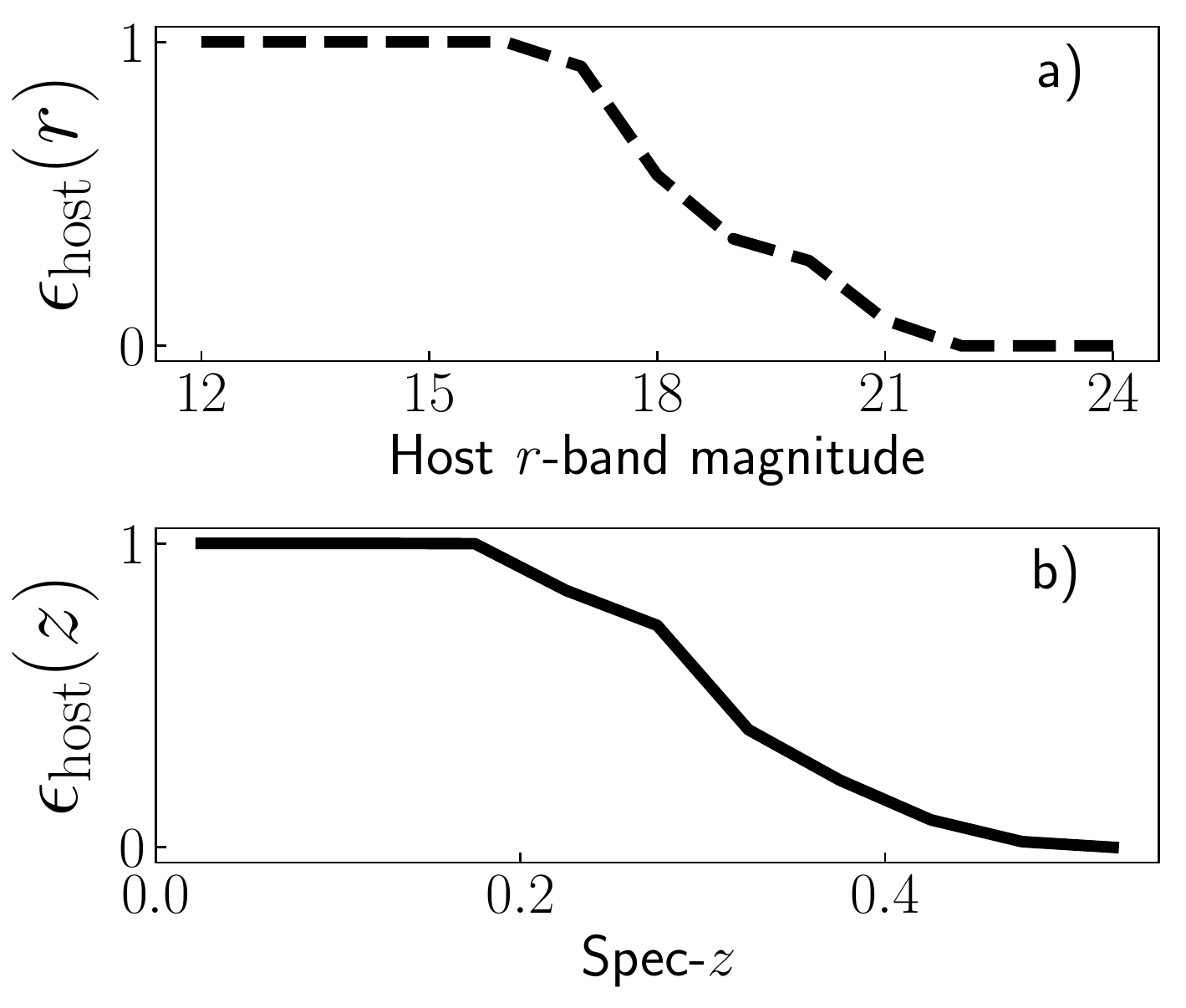}
\caption{The measured $r$-band host magnitude efficiency is shown in black dashed line and the measured host spec-$z$ efficiency is shown in solid black. The spec-$z$ efficiency is roughly half at $z = 0.3$. The $r$-band host magnitude efficiency drops off at the faint end.}
\label{Efficiency}
\end{figure}

We evaluated three different HOSTLIBs to use in our simulations. The first two, the Advanced Camera for Surveys General Catalog (ACS-GC) and the Marenostrum Institut de Ciències de l'Espai Simulations Catalogue (MICECAT), were used in \cite{Gupta16}.
For these two HOSTLIBs, the simulated $\SNhostas$ distribution does not match the SDSS data. 

Therefore, we created a third library by compiling observed galaxies within Stripe 82 from the SDSS DR14 data release. To maximise completeness and exclude spectroscopic selection effects, we selected host galaxies with a photometric redshift. The HOSTLIB includes Sersic profile information calculated from the Stokes values. These profiles are used in simulations to place supernovae near a galaxy and to model Poisson noise from the host galaxy. The SNANA simulation only calculates the smallest $d_{\textrm{\tiny{DLR}}}$ value for each SN, so the second smallest values were calculated separately from the HOSTLIB.

For data and simulation, Figure \ref{DLR-3fig} compares distributions of $\SNhostas$, DLR, and $d_{\textrm{\tiny{DLR}}}$. Each distribution is shown separately for the smallest and second smallest $d_{\textrm{\tiny{DLR}}}$ value.
Also shown is the ratio of smallest to second smallest $d_{\textrm{\tiny{DLR}}}$ values (\DDLRratio). We find good agreement in all distributions. Note that a HOSTLIB with too-large separations between galaxies can result in good data/sim agreement for the smallest $d_{\textrm{\tiny{DLR}}}$ in Figures \ref{DLR-3fig}a, b, and c, but would result in poor agreement for the second-smallest $d_{\textrm{\tiny{DLR}}}$, and also under-predict mis-associations. The good agreement for the second smallest $d_{\textrm{\tiny{DLR}}}$ and \DDLRratio~is therefore an important metric for reliably predicting the mis-association rate. We also show in Figure \ref{simcomphist}e that there is good agreement in the $r$-band magnitude of the host-galaxy distribution between data and simulations.

\begin{figure}[h]
\includegraphics[width=9cm]{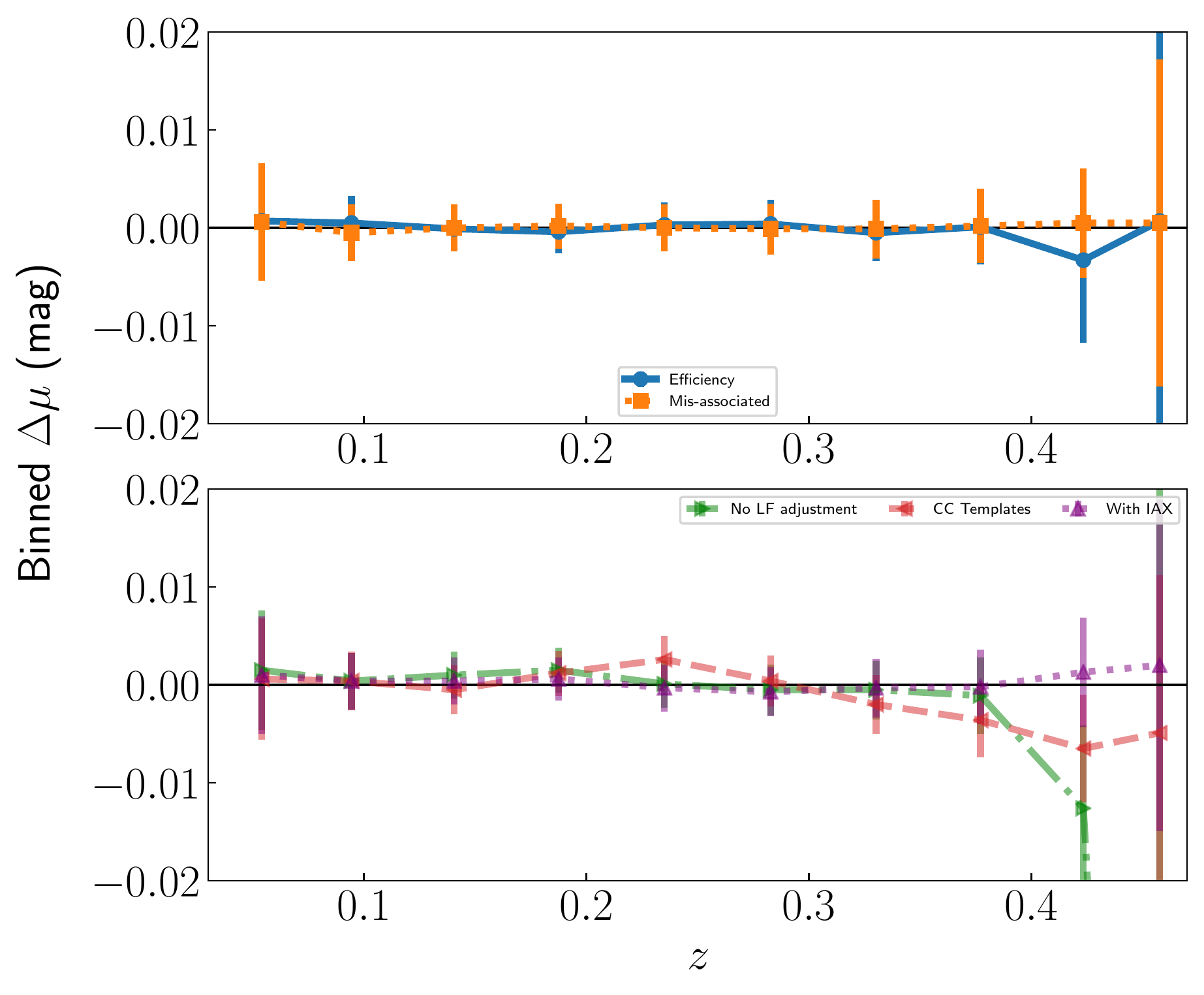}
\caption{The difference between binned distance modulus residuals for each systematic are shown here, for simulated supernovae. The top panel shows the effect of host galaxy selection changes, the bottom shows the different CC options.}
\label{M0Diff}
\end{figure}

Past studies have shown that there is a correlation between the stretch-and-colour-corrected luminosity of SNIa and the host galaxy stellar mass ($\stellarm$). This effect has been found in the Supernova Legacy Survey (SNLS; \citealp{Sullivan10}), the SDSS sample \citep{Lampeitl10, Wolf16, Hayden12, Gupta11}, and the PS1 sample (S18). In this SDSS analysis, we simulate these correlations to estimate biases arising from our spectroscopic galaxy selection. For every galaxy in our HOSTLIB, we calculate $\stellarm$ using the methodology from \cite{Taylor11}, \begin{equation}
    \stellarm = 1.15 + 0.7 \times ( g - i - 0.4 \times (i - \mu_{calc})),
    \label{Taylor}
\end{equation}
where \textit{g} and \textit{i} are the host galaxy magnitudes in the respective band, and $\mu_{\rm calc}$ is the calculated distance modulus using the galaxy redshift and same $\Lambda$CDM cosmology parameters as in the simulations. Equation \ref{Taylor} is used to calculate masses for both the data and the HOSTLIB. A comparison of the host-mass distribution between data and simulations is shown in Figure \ref{simcomphist}f; while the agreement is not as good as for the host-galaxy magnitude comparison and could use further study, it is sufficient to assess systematics.

We introduce a $-$0.025 mag correction to the luminosity of supernovae in galaxies with stellar mass \(M_\odot\) $>$ $10^{10}$,  and a $+$0.025 mag correction to those with \(M_\odot\) $<$ $10^{10}$. We do not include additional correlations of SNIa properties ($c$ and $x_1$) with host galaxy properties, these correlations are discussed in \cite{Smith14}.

Using simulations generated with input from our HOSTLIB, we evaluate the selection efficiency of our host galaxy. In our Fiduciary analysis, we define the efficiency, $\effHost$, to be a function of host galaxy $r$-band magnitude as follows,
\begin{equation}
       \effHost \equiv N_{\rm data}(r) / N_{\rm sim}(r)
       \label{hostr}
     \end{equation}
where $N_{\rm data}(r)$ is the number of SNe in each host galaxy $r$-band magnitude bin for data, and $N_{\rm sim}(r)$ is the number of SNe in each host galaxy $r$-band magnitude bin for a simulation with $\effHost = 1$. We scale $\effHost$ so that the maximum is 1 (Figure \ref{Efficiency}a).

For a systematic test, we follow \cite{Jones17} and parameterize the selection function to depend on host spec-$z$ (Figure \ref{Efficiency}b),

\begin{equation}
       \effz \equiv N_{\rm data}(z) / N_{\rm sim}(z)
     \end{equation}
where $z$ is the host galaxy spec-$z$, and $N_{\rm data}(z)$ and $N_{\rm sim}(z)$ are defined as in equation \ref{hostr}, but using $z$ bins.

\subsection{Core Collapse Simulations}
An important systematic in cosmological analyses of photometric samples is the collection of CC models used to simulate training samples for classifiers. The most recent study on this systematic has been done by \cite{Jones17}, which used a compendium of publicly available CC templates and adjusted the luminosity functions of the library to match the Hubble residual tail region after selection cuts. Light-curve templates of SNII were adjusted by 1.1 mag to be more luminous and SNIb/c were similarly adjusted by 1.2 mag. 

Since \cite{Jones17}, the PLAsTiCC library (\citealp{Kessler19, plasticc_models}) was released, which has enhanced previous CC template libraries.  Compared with \cite{Jones17}, here we include MOSFiT (Modular Open-Source Fitter for Transients) for SNIbc \citep{SNIbc_1, SNIbc_2, SNIbc_3, SNIbc_4}, NMF (Nonnegative Matrix Factorization) for SNII \citep{SNII_1, SNII_2, SNII_3, SNII_4}, SNIax \citep{SNIax_1}, and SNIa-91bg. We included an additional 0.9 mag smear for SNII-NMF as discussed in \cite{Kessler19}. Our Fiduciary analysis includes SNII-NMF, SNIbc-MOSFiT, and SNIa-91bg. As we will discuss in the next section, SNIax were excluded from both the simulated training and data samples.

\subsection{SNIax Simulations}\label{sec:level4:subsec:3}
The PLAsTiCC models include SNIax, which typically have lower luminosity, lower ejecta velocity, and greater variation in photometric parameters than their SNIa counterparts. The SED model used in PLAsTiCC is based on the real SNIax, 2005hk. The SED model was augmented with other spectra and the luminosity function was inferred from the sample studied in \cite{Jha17}. Light-curves are generated to match the absolute magnitude ($M_V$), rise time ($t_{\rm{rise}}$), and decline rate in the $B$ and $R$ bands ($\Delta m_{15}(B)$ and $\Delta m_{15}(R)$) detailed in \cite{Stritzinger15} and \cite{Magee16}.

\subsection{Simulation Analysis}\label{sec:4:level:4}

We apply the analysis (Section 3) to our simulated data sample, and fit for nuisance parameters $\alpha$, $\beta$, $\gamma$, and cosmological parameter $w$. The recovered values for these parameters in our Fiduciary analysis are consistent with their input values of 0.14, 3.2, 0.05, and $-1$, respectively. More precise validation tests with simulations are described in Section 6 of \cite{Brout18SYS}.


\section{Results}\label{sec:level5}

Here we assess the impact on our cosmological measurements of the systematics studied in this analysis, such as different CC templates used in classifier training sets, the frequency of host galaxy mis-association, and the modeling of selection efficiency. A summary of the various cosmological biases from these uncertainties is presented in Table \ref{tab:W-values}. The mean bias in $w$ is determined with the 40 simulated subsamples for each systematic listed in Table \ref{tab:W-values}, which has previously been discussed individually. In addition, the error on the mean and the scatter, or robust standard deviation, have also been calculated for the simulations. We define \begin{equation}
    \Delta w \equiv w_{Fid} - w_{sys}
\label{DeltaW}
\end{equation}
as the $w$-bias. For data, $w_{Fid}$ is the measured $w$ value for the Fiduciary analysis. For simulations, $w_{Fid}$ is the mean measured $w$ of our subsamples. The bias in recovered distances as a function of redshift due to each systematic uncertainty is shown in Figure \ref{M0Diff}. We define the statistical error ($w_{\rm stat}$) as 
\begin{equation}
    w_{\rm stat} = w_{\rm RMS}/\sqrt{N_{\rm sub}}
\end{equation}

where $w_{\rm RMS}$ is the RMS of $\Delta w$ and $N_{\rm sub}$ is the number of subsamples. With 40 subsamples, the statistical error in our mean $\Delta w$ is below 0.004, which is sufficiently small for this analysis. 

Host galaxy mis-association and shifts in the CC luminosity function result in a $w$-bias that can be corrected, and the resulting systematic uncertainty is typically smaller than the correction. Here we use the size of each correction as a systematic uncertainty, and in future work will evaluate the reliability of these corrections along with the associated systematic uncertainties.

\subsection{Galaxy Association and Mis-association} \label{sec:level5:subsec:3}
 
Figure \ref{DLR-3fig} shows the properties used to validate the simulations from which the mis-association rate was determined. Comparing each true host galaxy in our simulation to the $d_{\textrm{\tiny{DLR}}}$-selected galaxy, we determine the host galaxy mis-association to be 0.6\%.  
 
\begin{figure}[h]
\includegraphics[width=9cm]{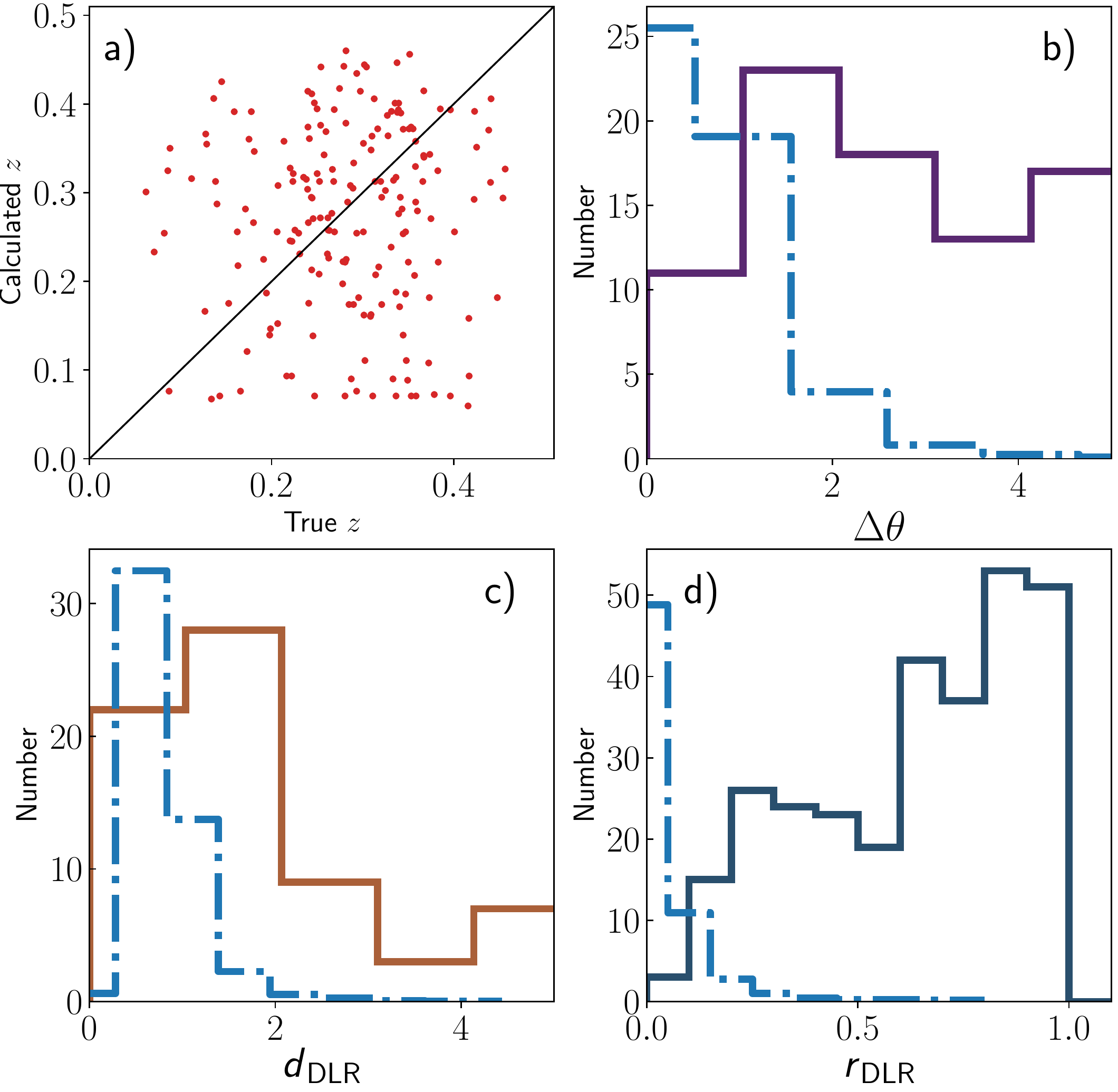}
\caption{For simulated supernovae with mis-associated hosts, Panel a) shows true values of redshift versus the calculated values, with the black line showing correct association for reference. The solid histograms in panels b), c), and d) show distributions only for the mis-associated supernovae, the dash-dotted histogram for all supernovae (normalised to mis-associated population). When the galaxy is mis-associated, we attribute SNe to higher-redshift hosts 56\% of the time and lower-redshift hosts 44\% of the time.}
\label{DDLR-Ratio}
\end{figure}

Figure \ref{DDLR-Ratio} shows the effect of mis-associated hosts on redshifts (\ref{DDLR-Ratio}a), as well as the distributions of $\SNhostas$ (\ref{DDLR-Ratio}b) and $d_{\textrm{\tiny{DLR}}}$ (\ref{DDLR-Ratio}c) for those mis-associated hosts; figure \ref{DDLR-Ratio}d is a histogram of \DDLRratio~for mis-associated SN. In each panel, the distribution for mis-associated hosts is much broader compared to correctly identified hosts.

We find that the recovery of $w$ is biased by $\Delta w =$ 0.0007 by this mis-association, which is consistent with 0.

\subsection{Impact of Host Galaxy Selection Efficiency}

We evaluate the bias on $w$ due to our host galaxy selection by comparing our recovered distances using
$\effHost$ to those using $\effz$. The impact on the binned distances is shown in Figure \ref{M0Diff}. For most of the redshift range, the impact is less than 2 milli-mags and the only significant impact is at the higher end of our redshift range. This produces a $w$-bias of $\Delta w = $ \MagEffp{}, significantly smaller than the statistical uncertainty. The RMS around this value is \MagEffsig{}.

\begin{table}[]
\caption{$w$ Differences for Systematic Tests}
\scalebox{.9}{
\begin{tabular}{l|l|l|l|l}
Systematic Test &  &  &  &  \\
(Host) & $\Delta w_{\rm sim}$\footnote{Mean $\Delta w$ of the 40 simulated subsamples.} & $w_{\rm RMS}$\footnote{The Root Mean Square of the simulated subsamples.} & $\Delta w_{\rm data}$\footnote{The $\Delta w$ measured by the data.} & $N_{\sigma}$\footnotemark[4] \\
\hline
Mis-associated Host & \WrongHost{} &  \WrongHostS{} & N/A & N/A\\
Host Efficiency &  \MagEff{} & \MagEffS{} & \MagEffData{} & \MagEffsig{} \\
 & & & & \\
 & & & & \\
Systematic Test &  &  &  &  \\
(Contamination) & $\Delta w_{\rm sim}$ & $w_{\rm RMS}$ & $\Delta w_{\rm data}$ & $N_{\sigma}$ \\
\hline
No LF adjustment & \PBA{} & \PBAS{} & \PBAData{}  & \PBAsig{} \\
Choice of CC model & \DJones{} & \DJonesS{} & \DJonesData{} & \DJonessig{} \\
Include Iax & \IAX{} & \IAXS & \IAXData{} & \IAXsig{}

\footnotetext[4]{The number of standard deviations of $\Delta w$ for data away from $\Delta w$ for sim, defined in equation \ref{Sigma}.}
\end{tabular}}
\label{tab:W-values}
\end{table}

\subsection{Impact of Core Collapse Templates}

\begin{figure*}[t]
\includegraphics[width=18cm]{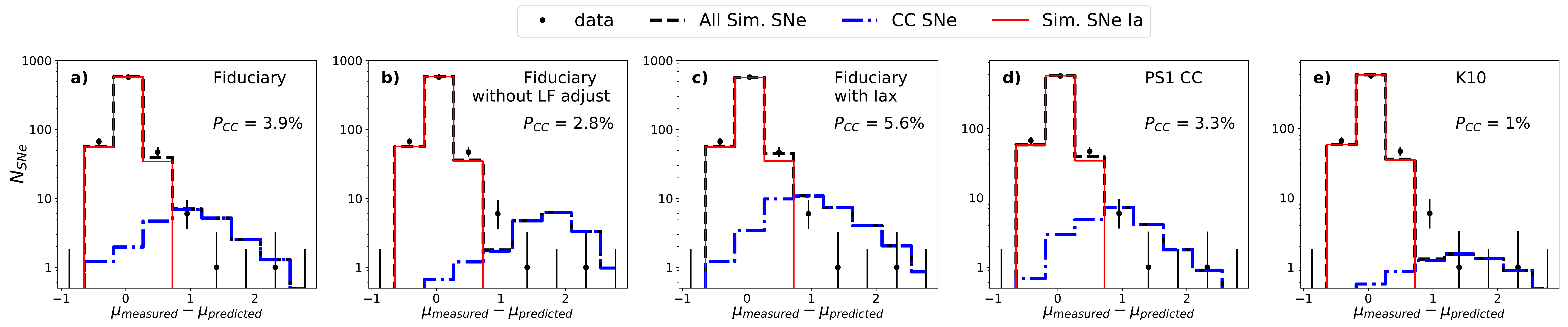}
\caption{The Hubble residual distribution for data, and for simulations using different CC models as indicated on each panel. $\mu_{measured}$ is the measured Tripp distance modulus and $\mu_{predicted}$ is the predicted distance modulus from $\Lambda$CDM cosmology. The training sample is the Fiduciary CC model in all cases. Data is shown with points, simulated SNIa are shown with a red histogram, and simulated CC SNe are shown with a blue dashed histogram. The combined simulated distribution of CC and Ia SNe is shown with a black dashed histogram. The non-SNIa contamination ($P_{CC}$) is shown for each panel.}
\vspace{30pt}%
\label{Hubble-Residuals}
\end{figure*}

Figure \ref{Hubble-Residuals} shows the Hubble residual distributions using five different CC models in the analysis, and each model is indicated in the panel. Figure \ref{Hubble-Residuals} contains our Fiduciary case, where CC templates used in the simulations are from: the PLAsTiCC models with an adjusted luminosity function and without SNIax; the PLAsTiCC model with neither adjusted luminosity function nor SNIax; the PLAsTiCC model with adjusted luminosity function and SNIax; the CC templates from \cite{Kessler10} without luminosity adjustments (K10); and the CC templates from \cite{Jones18} with adjustments. The smallest contamination ($P_{CC}$) is \CCKTen{}\% for K10, and the highest is \CCIAX{}\% for the Fiduciary with Iax analysis. Still, we see that the contamination in the data for the positive tail of the Hubble residual distribution is better predicted in some cases than others. We find that the PLAsTiCC and K10 models do not match the data well in this region, confirming the need for luminosity function corrections to match the high Hubble residual tail suggested in \cite{Jones18}.

For the case of Fiduciary with SNIax, the SNIax make up 40\% of the CC contamination. While the SNIax luminosity distribution is fainter than SNIa, we also find that the $x_1$ and $c$ values satisfy the selection requirements (see section 2.2), and therefore the NN classification poorly separates SNIa from SNIax. \cite{Jones17} do not include SNIax in the contamination library due to the expectation that they are too red ($c$ $>$ 0.3), and because the SNIax model was not available. As discussed in section \ref{sec:level4:subsec:3}, only a single SN was used to generate the SNIax model. However, if we perform SALT2 light-curve fits on the four known SNIax in SDSS (including 2005hk), we find that all the SNIax have fitted colour values $c$ $>$ 0.3 and thus fail our selection requirements. Therefore, these light-curve fits suggest that the SNIax contamination is overestimated and further study is needed. 

From simulations, the impact on $w$ due to the systematics related to core collapse libraries is given in Table \ref{tab:W-values}. We find the mean bias due to the CC systematics is $\Delta w < |0.01|$ with a statistical uncertainty of $\sim0.001$. The RMS in $w$ from the simulations due to the systematics is 0.01-0.02.

\subsection{Data and Simulation Comparison}
In the last two columns of Table \ref{tab:W-values}, we show the impact on $w$ of the systematics studied in this analysis for the real data sample.  This is shown for all the systematics except for the mis-associated host, as there we can not apply the same technique on the data as we did for the simulations.  To assess whether the changes seen for the data sample are consistent with predictions from the simulations, we define the number of standard deviations ($N_{\sigma}$) for each systematic as 

\begin{equation}
    N_{\sigma} = (\Delta w_{\rm sim} - \Delta w_{\rm data}) / w_{\rm RMS}
    \label{Sigma}
\end{equation}
where $\Delta w_{\rm sim}$ is the $\Delta w$ recovered in simulations, $\Delta w_{\rm data}$ is the $\Delta w$ recovered from the data, and $w_{\rm RMS}$ is the RMS of the $\Delta w$ recovered from simulations. We find that the highest deviation compared to the simulations is seen for the `No LF adjustment' systematic at $2.1\sigma$.  All other deviations near or below $<1\sigma$. Therefore, we conclude that the impacts of the systematics seen in the simulations are consistent with those seen for the data.

\section{Discussion and Conclusions}\label{sec:level6}

In this paper, we have presented new methodologies for two systematic uncertainty contributions unique to analyses of cosmology with photometric SNIa samples: 1) host galaxy mis-association and selection efficiency, and 2) core collapse training library. For classifier training and bias corrections, we generated realistic simulations of supernova (SNIa and CC) and host galaxies. We validated these simulations with a wide range of diagnostics. We find the host galaxy mis-association rate to be 0.6\%, resulting in a $w$-bias of $\Delta w = $ 0.0007. We expect the mis-association rate, and hence the distribution of mis-associated redshifts, to change with the redshift range of a survey. If the impact of this systematic increases in future analyses, more rigorous bias correction simulations and possibly new analysis methods such as $z$-BEAMS \citep{z-BEAMS} may be necessary.

The galaxy selection efficiency contributes a $w$-bias of $\Delta w = $ \MagEffp{}. For the first time, the PLAsTiCC library has been used for assessing systematic uncertainties in a cosmological analysis. We confirm the \cite{Jones18} finding that CC luminosity function adjustments are needed to more accurately predict the Hubble residual tail (see Figure \ref{Hubble-Residuals}). We find that ignoring the CC luminosity function shift results in a $w$-bias of $\Delta w = $ $-0.0109 \pm 0.0003$.

The scale of these systematics is similar to that found in \cite{Jones18}, given similar priors, for the contamination systematics, though they do not explicitly include systematics for galaxy mis-association or the efficiency of host galaxy follow-up. The systematic shifts in the data are well predicted by the simulations as shown in Table \ref{tab:W-values}.

The total statistical uncertainty on $w$ from a cosmological fit to the SDSS sample with the same priors as discussed in section 4 is 0.1, larger that the systematics investigated here. For larger samples, the statistical uncertainty will be smaller and systematic uncertainties of this size will be more significant. Here we have developed a framework that can be used  to evaluate these systematic uncertainties in future analyses of photometric samples.

\appendix \label{sec:appendix}

\subsection{Discussion of $d_{\textrm{\tiny{DLR}}}$ values}

We start with the radial equation of an ellipse as measured from the center, with semi-major and semi-minor axes $a$ and $b$ and orientation angle $\theta$:

\begin{equation}
    r(\theta) = \frac{ab}{\sqrt{(a \textrm{sin}\theta)^2 + (b \textrm{cos}\theta)^2}}
    \label{DLR-EQ}
\end{equation}

We define the supernova angle as the angular difference between a line that goes through the supernova position and galaxy center and a line that passes through North and the galaxy center. Combined with the orientation of the galaxy as given by the galaxy position angle we define $\theta$ by subtracting the supernova angle from the position angle of the galaxy. With $a$, $b$, and $\theta$, the directional light radius (DLR) is defined from equation \ref{DLR-EQ} as the effective radius of the galaxy at angle $\theta$.

The position angle of the galaxy can be found using the Stokes parameters Q and U given in the SDSS DR14 data release. DR14 uses a slightly unconventional notation for U; this is corrected with a factor of 2 in the position angle.

Since these parameters are not fits to a model, but rather based on pixel data, they are more robust for fainter galaxies. The position angle $\phi$ can be expressed as 
\begin{equation}
    \phi = \frac{1}{2} \textrm{arctan}(\frac{2U}{Q})
    \label{Position-Angle}
\end{equation}
The ratio of the semi-major and minor axes can also be computed with the Stokes parameters. Defining $\kappa \equiv Q^2 + U^2$, the ratio \textit{a/b} is then expressed as 

\begin{equation}
    \frac{a}{b} = \frac{1+ \kappa + 2\sqrt{\kappa}}{1 - \kappa}
    \label{ab}
\end{equation}

Following S14, we set $a$ equal to the Petrosian half light radius within the $r$-band and $b$ is determined using equation \ref{ab}.
Finally, we define a distance weighted $d_{\textrm{\tiny{DLR}}}$

\begin{equation}
    d_{\textrm{DLR}} = \frac{\textrm{Angular separation}}{r(\theta)}
    \label{DDLR}
\end{equation}

It is important to note that DLR and $d_{\textrm{\tiny{DLR}}}$ are survey dependent quantities and are not easily comparable across surveys. Of particular note are magnitude cutoffs. Establishing a magnitude limit does allow for fine tuned control of $\SNhostas$ density, but does not account for apparent ellipticity. At higher magnitudes, the apparent ellipticity as measured by the Stokes parameters begins to increase. At fainter brightness, noise begins to dominate the signal and leads to unrealistic ellipticity measurements. But differences in magnitude limits for different surveys, combined with those in image processing, can alter the apparent size of a galaxy. Self-consistent DLR measurements within the survey are more accurate than solely $\SNhostas$ determinations, but cross comparison would not be effective.

\bibliographystyle{apj}
\bibliography{research2.bib}

\end{document}